\documentclass{article}

\usepackage{arxiv}
\usepackage[utf8]{inputenc} \usepackage[T1]{fontenc}    \usepackage{hyperref}       \usepackage{url}            \usepackage{booktabs}       \usepackage{amsfonts}       \usepackage{nicefrac}       \usepackage{microtype}      \usepackage{lipsum}  \usepackage{listings}
\usepackage{graphicx}
\usepackage{subcaption}
\usepackage{xcolor}
\title{VisJSClassificator - Manual Visual Collaborative Classification Graph-based Tool}
\author{
  Vincent~Falconieri\\
  CIRCL\\
  Luxembourg \\
}
\begin{document}
\maketitle
\begin{abstract}
Analysts need to classify, search and correlate numerous images. Automatic classification tools improve the efficiency of such tasks. However, classified data is a prerequisite to develop these tools. Labelling tools are of great use in case of already known classes, but seemed limited for Open Set Classification. This paper presents a manual and collaborative classification tool, which uses graph representation.
\end{abstract}


\section{Introduction}
CERTs - as CIRCL - and security teams collect and process content such as images (at large from photos, screenshots of websites or screenshots of sandboxes).
In datasets become larger - e.g. on average 10000 screenshots of onion domains websites are scrapped each day in \textbf{AIL}\footnote{Analysis Information Leak framework - \href{https://github.com/CIRCL/AIL-framework}{github.com/CIRCL/AIL-framework}}\cite{mokaddemAILDesignImplementation2018}, an analysis tool of information leak - and analysts need to classify, search and correlate through all the images. \\
Automatic tools can help them in this task. Image matching and image classification tools are great allies to discharge analysts from this burdening task. However, datasets and their classification are decisive inputs to build these tools. We considered a context of Open Set Classification\cite{scheirerOpenSetRecognition2013a}, where many classes are known but many aren't. Labeling software were of limited interest in this context and in our case, as new classes appear during the labeling. \\
Therefore, we've built for this circumstance a manual, visual, collaborative and graph-based classification tool named \textbf{VisJS-Classificator}.
\subsection{Problem Statement}
Labeled datasets can be a prerequisite to consistently evaluate algorithms classification performances. If the dataset content is unknown before the labeling (which is the case of any scrapped data), one may need to read once the whole dataset to establish labels, to then read a second time the dataset to label it.
This method raises few issues. How to ensure the consistency between the label meaning established at the beginning and at labelling time ? How to ensure that when a label is chosen, already similar pictures have the same label ? etc.
Standard labeling tools, as \textbf{Dataturks}\footnote{OpenSource Data Annotation tool for teams - \href{https://github.com/DataTurks/DataTurks}{github.com/DataTurks/DataTurks}}, can't display all pictures already present in each cluster. Therefore, consistency between already labeled pictures and currently labeled picture cannot structurally be guaranteed. In this setting, labels have to be chosen prior to the labeling phase.
For our Open Set Classification settings, we choose to build a labelling tool which does not need prior labels list, and allow to see what is in which and every cluster at all time.
\section{Proposed Approach}
We have chosen a graph-based approach to meet our requirements : no dependency to a prior labels list and a continual visualisation of clusters content. 
\subsection{Creation of the tool}
Technically, we used \textbf{VisJS}\footnote{OpenSource Data Annotation tool for teams - \href{https://github.com/visjs}{github.com/visjs}} as a basis on which we built up VisJS-Classificator. VISJS was at first only a way to get a quick overview of a dataset and our algorithms outputs. After few modifications, our modified version of VISJS became a labeling tool and its output serves as input for evaluation algorithms.
Few main components were added to the original VisJS visualisation library : collaborative aspects, labeling actions, pictures display, export and import of data structures, ...
\subsubsection{Collaborative aspects}
We added a NodeJS server to VisJS Javascript visualisation client, as well as a \href{https://github.com/socketio/socket.io}{Socket.io} communication between client and server. Then clients and server can share the state of the graph in real time, allowing collaborative classification to be performed.
The position of each picture within the display is not shared, and so remains only on the client side. This design choice occurs to prevent a large number of positions to be shared between clients and servers, which could have decreased refresh performances without favouring classification easiness.
\begin{figure}[h!]
  \centering
  \includegraphics[width=0.6\linewidth]{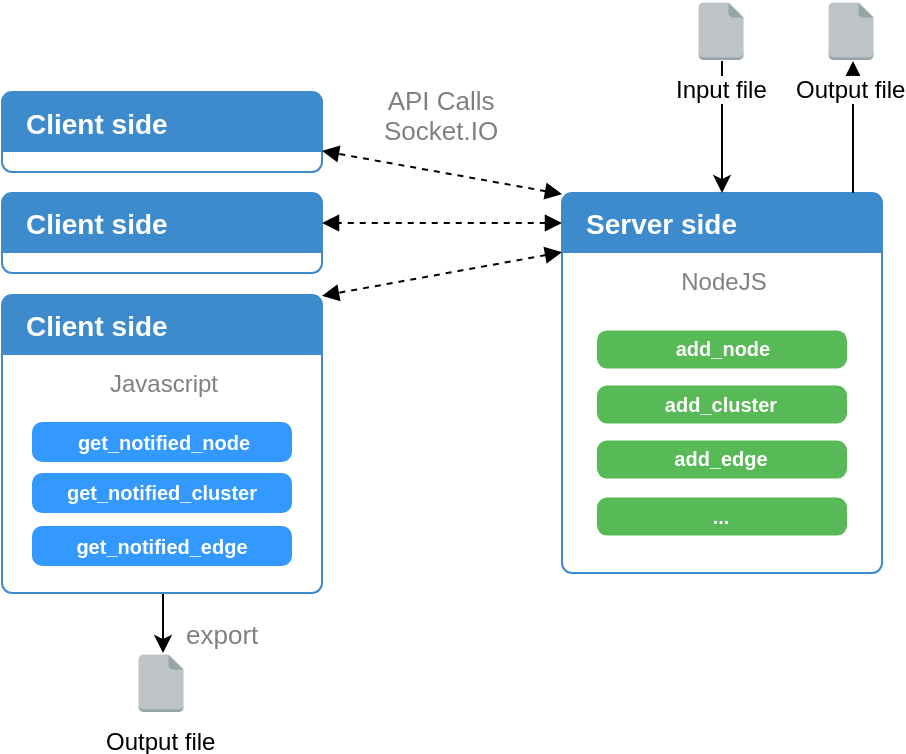}
  \caption{Architecture overview}
  \label{fig:visjs_overview}
\end{figure}
\clearpage
\pagebreak
\subsubsection{Labelling Actions}
Users can classify (cluster and name) pictures with a few actions.
\begin{itemize}
\item They can drag and drop pictures with a dragged left click, any picture or node.
\item They can select pictures with a dragged right click.
\item They can cluster pictures with a press on 'C'. This creates a new node called "anchor" which operate as a cluster representation. Each picture of the selection gets a new edge from itself to this new "anchor". This binds each picture with this particular cluster.
\item They can also rename a node (including an "anchor", and so a cluster) by selecting a node and press an "Edit Node" button on the user interface. This opens a dialog where the user can change the label of the node, and so, the label of a cluster.
\end{itemize}
\begin{figure}[h!]
  \centering
  \begin{subfigure}[b]{0.54\textwidth}
    \includegraphics[width=\linewidth]{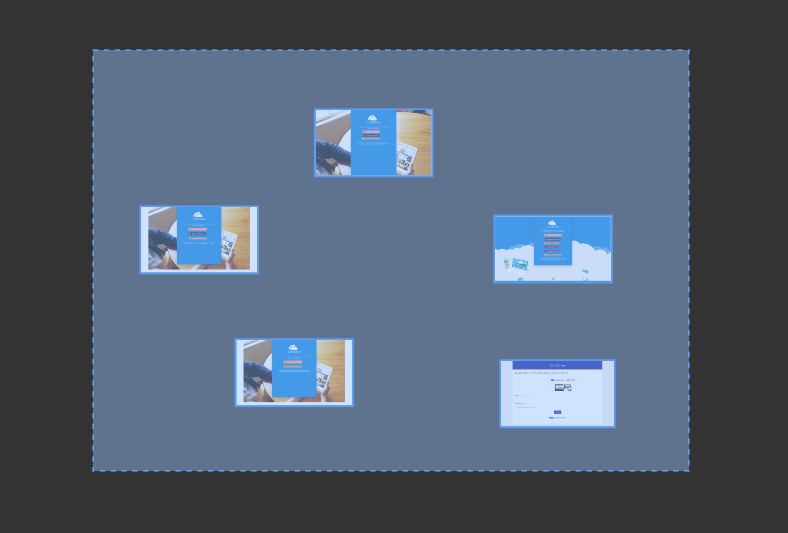}
    \caption{A selection over some pictures}
  \end{subfigure}
    \begin{subfigure}[b]{0.45\textwidth}
    \includegraphics[width=\linewidth]{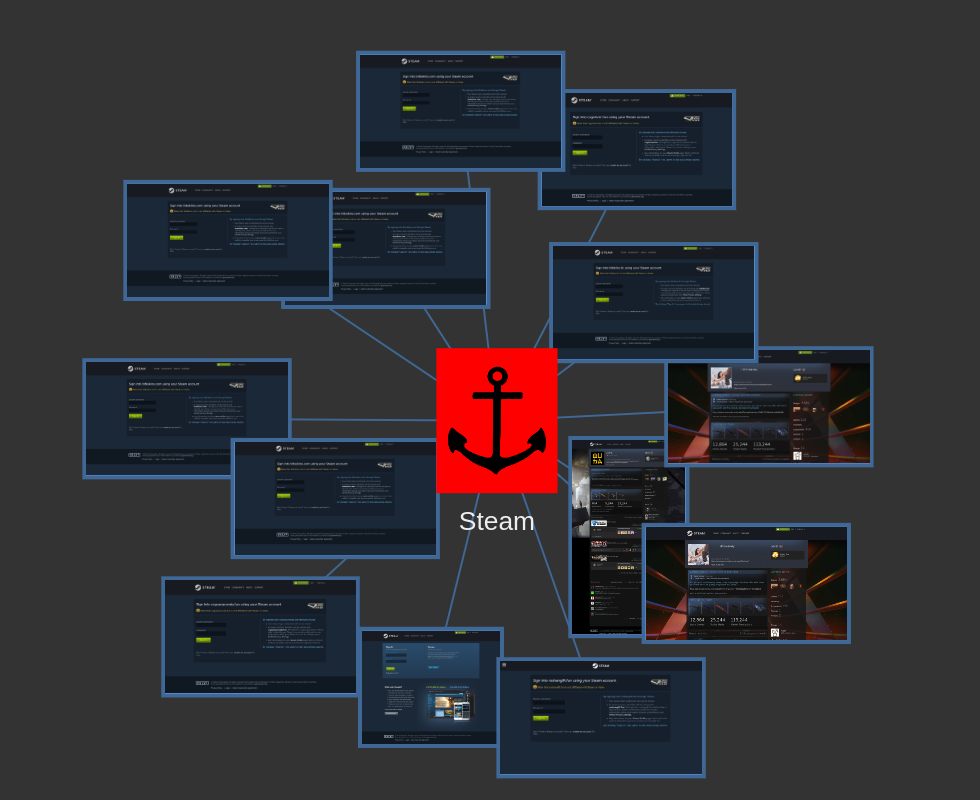}
    \caption{Clustered pictures}
  \end{subfigure}
  
  \begin{subfigure}[b]{0.42\textwidth}
    \includegraphics[width=\linewidth]{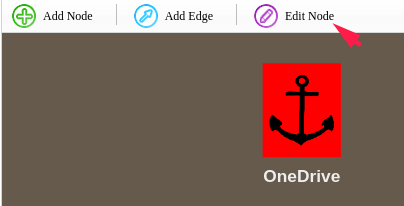}
    \caption{Renaming a node}
  \end{subfigure}
  \begin{subfigure}[b]{0.52\textwidth}
    \includegraphics[width=\linewidth]{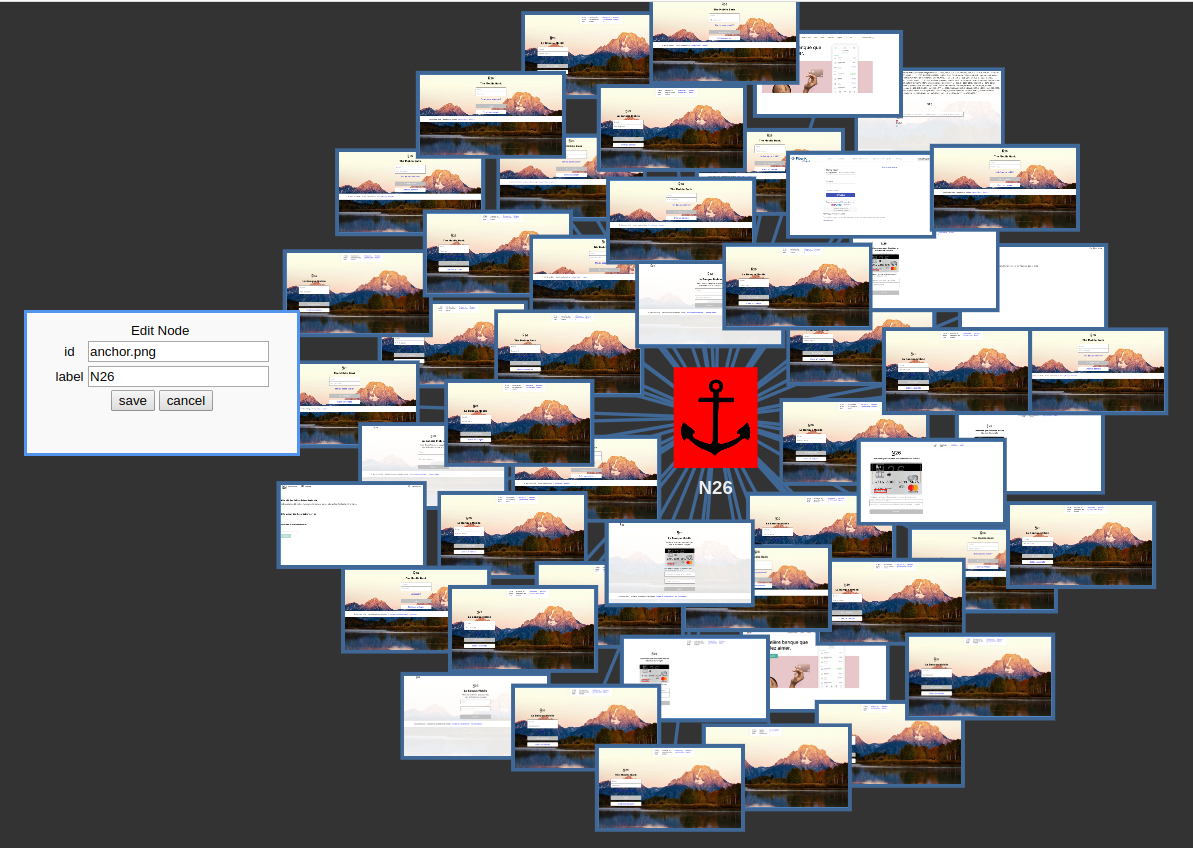}
    \caption{Windows to rename an anchor and so a cluster of pictures}
  \end{subfigure}
  \caption{Cluster creation steps : selection, clustering, renaming, confirmation}
  \label{fig:clusters-example}
\end{figure}
\clearpage
\pagebreak
\subsubsection{Scaling picture display}
\begin{figure}[h!]
  \centering
  \begin{subfigure}[b]{0.5\textwidth}
    \includegraphics[width=\linewidth]{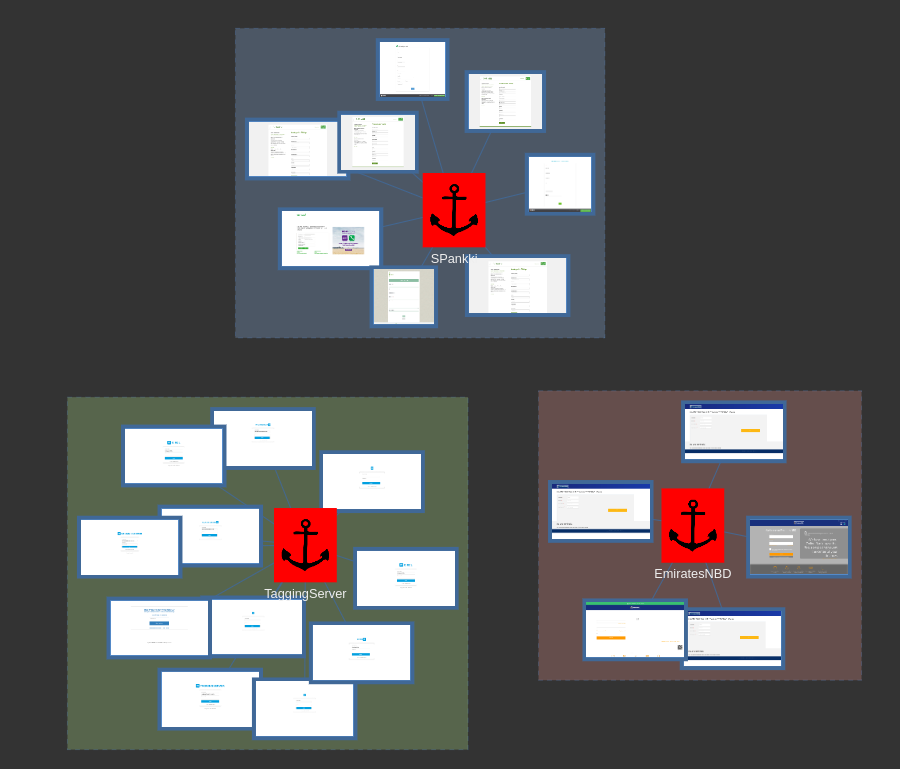}
    \caption{HighPerformance mode disabled : cluster boxes drawn, all pictures visible, ...}
      \label{fig:highperfdisabled}
  \end{subfigure}
    \begin{subfigure}[b]{0.49\textwidth}
    \includegraphics[width=\linewidth]{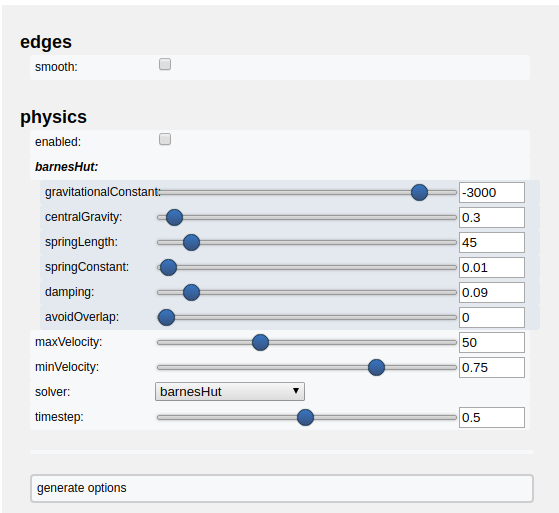}
    \caption{VISJS Classificator options.}
  \end{subfigure}

  \caption{Highperformance mode and options in VisJS-Classificator}
  \label{fig:interface-info}
\end{figure}

A lot of pictures are slowing down the visualisation of the graph.\\
This speed can be improved by not tracing cluster boxes around the cluster, by not showing full-size pictures but only down-scaled version of them, by not tracing all edges, by turning off the physics simulation of nodes, by fixing position of most of nodes, etc.\\
A still-in-development feature is a "High-performance" toggle button, which would activate or not all of these actions. Therefore, bigger graphs can still be manually clustered, as the high-performance mode allows to show more pictures at the same time with decent performance. See Figure \ref{fig:highperfdisabled} for an example without Performance mode activated. \\
So far, graphs up to 4000 pictures had been loaded with a decent frame rate (>10 im/sec) allowing classification, on a high-end laptop\footnote{Intel(R) Core(TM) i7-4500U CPU @ 1.80GHz , 12Gb RAM}.

On launch, pictures can be resized to any resolution, to lower the system requirements on the first launch (and potentially needed to events load the webpage). The burden is on the client side and not server side, as the main bottleneck is the visualisation of the graph and not its management.
\clearpage
\pagebreak
\subsubsection{Custom VisJS-Classificator Graph data-structure}
Current working graphs can be exported on the client side and on the server side. A previously saved graph can be loaded only on the server side to guarantee consistency and uniqueness of the graph shown to every client.
Our tool can export the current state of the graph and reload it later. It can also create a graph structure from nothing more than a folder of files. A JSON extract (Listing \ref{graph-struct}) presents the graph export of VISJS-Classificator.
\begin{lstlisting}[frame=single,caption=Graph structure,label=graph-struct]
  (...) 
  "clusters": [
    {
      "id": "a5e1baa2-aead-4164-9205-63f26f656d6f",
      "image": "anchor.png",
      "label": "BancoInter",
      "shape": "image",
      "group": "anchor",
      "members": [
        20,
        12,
        17,
        (... more members ...)
      ]
    }, 
    (... more clusters ...)
  ],
  "nodes": [
    {
      "id": 0,
      "image": "abashed-careless-ordinary-crew.png",
      "shape": "image"
    }, (... more nodes ...)
    {
      "id": 456,
      "image": "zonked-silent-snobbish-review.png",
      "shape": "image"
    }
  ],
  "edges": [
    {
      "to": "a5e1baa2-aead-4164-9205-63f26f656d6f",
      "from": 20
    }, 
    (... more edges / links ...)
  ]
\end{lstlisting}
\clearpage
\pagebreak
\section{Materials and Methods}
We evaluated the speed over two benchmarks. The first benchmark was conducted by 3 participants. Due to the length of the task, the second benchmark has only been performed with one person willing to spend that much time on the task. These results are not statistically representative but can still give insights about qualitative difference between a "\textit{standard labelling tool}" and "\textit{a graph-based labelling tool}". \\
The first setup uses 85 pictures randomly sampled from circl-phishing-dataset-01, which's a large 475 phishing pictures dataset.\\
The second setup uses the full 475 pictures of the same dataset.

\subsection{Test Methodology}
Participants who were monitored during their labelling had no prior experience about the labelling tool, nor the "\textit{standard}" one nor the "\textit{graph-based}" one. This prevents any experience bias.
Labels were not provided, as in the real situation, and had to be defined by the participant during the experiment. Labels were added directly by the participant and were counted in the completion time.\\
Times are recorded from the first "click" on the interface, after the full loading of the application, until the last click on the interface, when users had labeled all pictures at least with one tag. 

\subsection{Test environment}
Screenshots of the test environment are presented in Figure \ref{fig:user-interfaces}, page \pageref{fig:user-interfaces}.

\begin{figure}[h!]
  \centering
  \begin{subfigure}[b]{0.6\textwidth}
    \includegraphics[width=\linewidth]{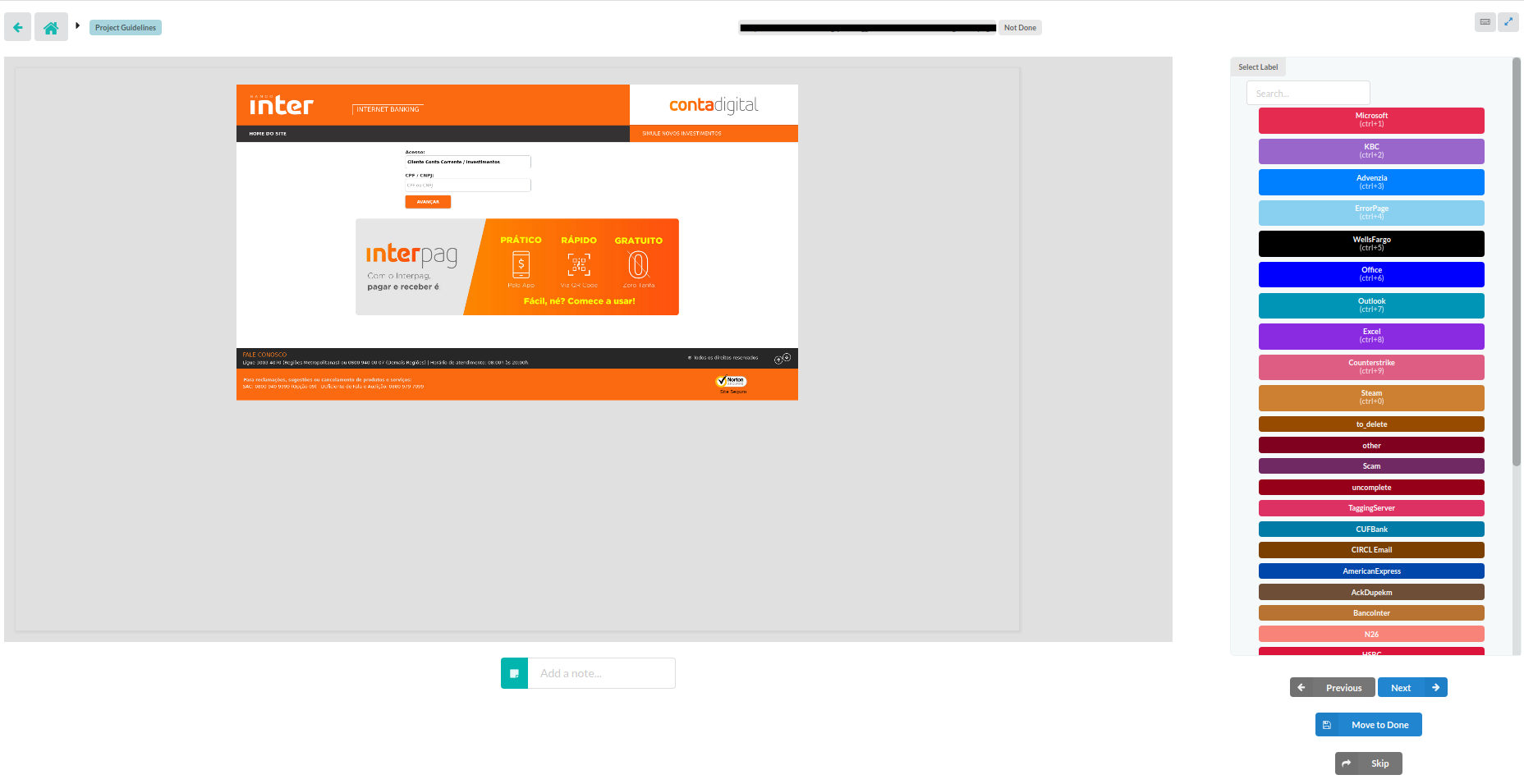}
    \caption{DataTurks Labelling interface. On the right-hand side is the picture to be labelled, on the right-hand side are labels to choose from with a search field. Labels and actions (left/right/accept pictures) have keyboard shortcuts.}
    \label{fig:labelingInterfaceDataTurks}
   \end{subfigure}
  
    \begin{subfigure}[b]{0.6\textwidth}
    \includegraphics[width=\linewidth]{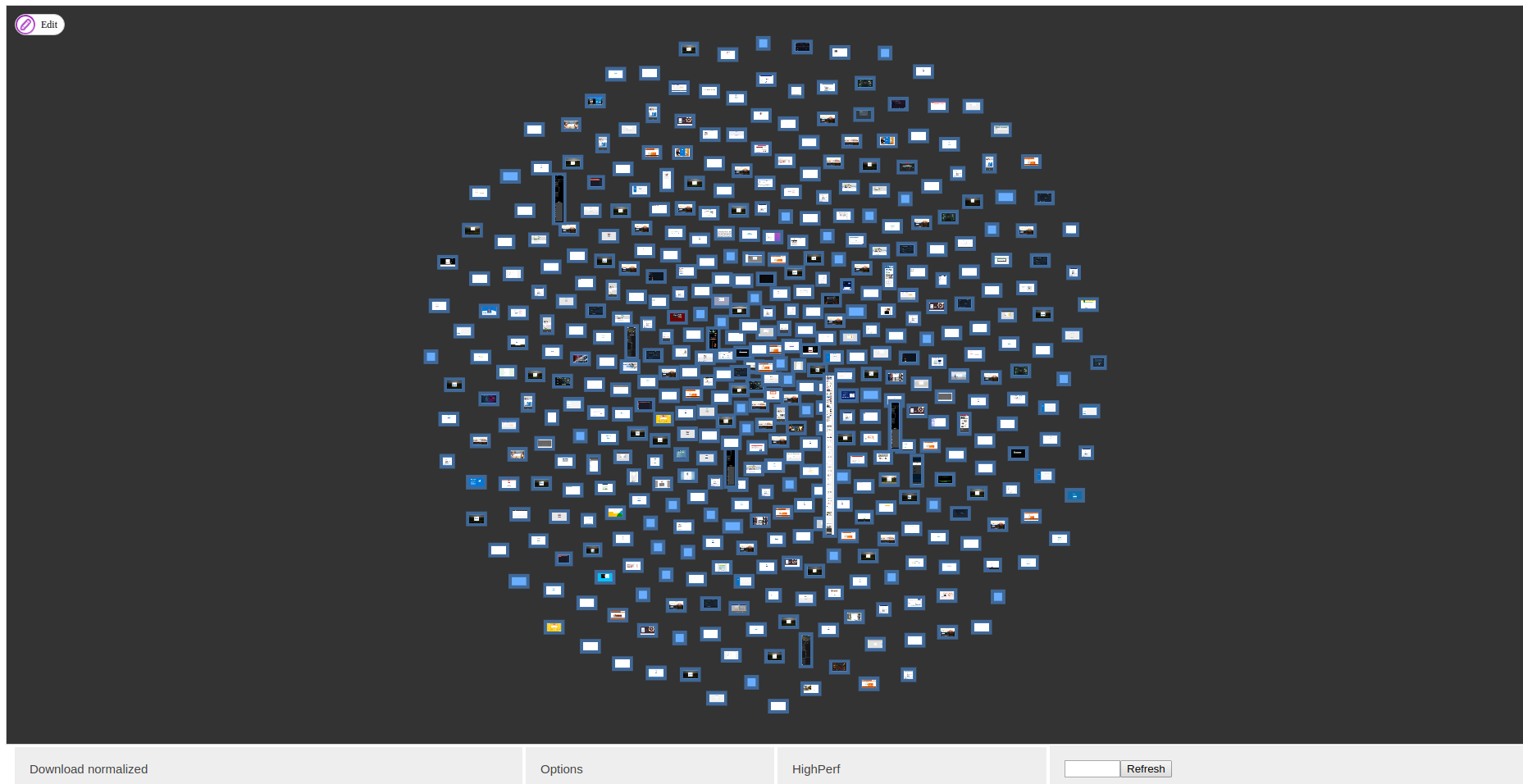}
    \caption{VISJS Classificator labelling interface. On the middle of the screen are pictures to be labelled, at bottom are buttons to modify graph (physics ...). Most actions are accessible with shortcuts, e.g. left click (for selection) and 'C' key to cluster currently selected pictures.}
    \label{fig:VisJSTriageInterface}
  \end{subfigure}
  
  \caption{User interfaces}
  \label{fig:user-interfaces}
\end{figure}

\subsection{Results}

\subsubsection{85-pictures experiment}

\begin{table}[h!]
\centering
\begin{tabular}{cccl}
\textbf{Participant} & \textbf{Time Standard labelling tool} & \textbf{Time graph based labelling tool} & \\
1             &  6 minutes (\_+3+3)                  & 8 minutes                   &\\
2             & 19 minutes (6+7+6)                   & 42 minutes (first)          &\\
3             & 35 minutes (13+14+8) (first)         & 56 minutes				      \\
\end{tabular}
\caption{First experience, 85 pictures to label}
\label{Tab:first-exp}
\end{table}

For participant 1, labels were provided in both cases. Displayed time is only classification time. Dataset was already known to the participant.\\
For participant 2, 19 minutes of classification were decomposed in 6 minutes (labels finding) + 7 minutes (50\% labeled) + 6 minutes (100\% labeled).\\
For participant 3, 35 minutes of classification were decomposed in 13 minutes (labels finding) + 14 minutes (50\% labeled) + 8 minutes (100\% labeled).

Findings are : 
\begin{itemize}
\item Improvement in speed is clear, even during the classification of one unique dataset. Participants got used to the tool and the labels they choose. Their efficiency greatly increase as they are labeling.
\item Participants using VisJS Classificator are slower than standard labelling tool, and the difference is greater if the experience in labeling tool and the dataset is lower.
\item Minors improvements to VisJS Classificator may greatly impact participant efficiency : displaying a thumbnail of a picture on mouse hover (to reduce zoom in/out), display bigger cluster names, etc. These constitutes futures direction of development.
\end{itemize}

\subsubsection{470-pictures experiment}

\begin{table}[h!]
\centering
\begin{tabular}{cccl}
\textbf{Participant} & \textbf{Time Standard labelling tool} & \textbf{Time graph based labelling tool} &   \\
1             &               49 minutes                       &                    64 minutes                     &   \\
\end{tabular}
\caption{Second experience, 475 pictures to label}
\label{Tab:second-exp}
\end{table}

First case is a normal labelling classification of 470 pictures to 56 different final labels and not mutually exclusive, defined during the labelling. 49 minutes were needed, which is a 6.25 seconds per picture speed. This time includes labels adding when few pictures with the same "look" were seen. 

\begin{itemize}
\item We have to highlight that the relative quality of clustering/labelling was lower with dataturks than with visjs. Mainly, users tend to be less consistent when they don't see other pictures labelled with one given label. Similar pictures which could match three different labels were labelled with subsets of these three labels.
\item There is an aversion to label creation. One of the negative consequences for the labeller is that he would need to go back to already labelled data to double-check if the picture would fit in this new label. 
\item One other negative consequence is that the creation of a label is sometimes not seen as "worth-it". If very few pictures (upon discussion <= 2) have to be labelled, users prefer to label them as "other" instead of creating a new label.
\end{itemize}

Second case is a graph-based labelling classification of 470 pictures to 42 different final and mutually exclusive labels, defined during the labelling. 64 minutes were necessary, which is a 8 seconds per picture speed. This time includes labels addition when few pictures with the same "look" were seen.

\begin{itemize}
\item We have to highlight that the time to find biggest clusters is far lowered by visjs settings : similar pictures are easy to spot, and within the first 10 minutes, major clusters were already found.
\item Time was mostly lost on small clusters (<= 2 pictures) where it was necessary to look more in details to match pictures.
\end{itemize}

\begin{figure}[h!]
  \centering
  \begin{subfigure}[b]{\textwidth}
    \includegraphics[width=\linewidth]{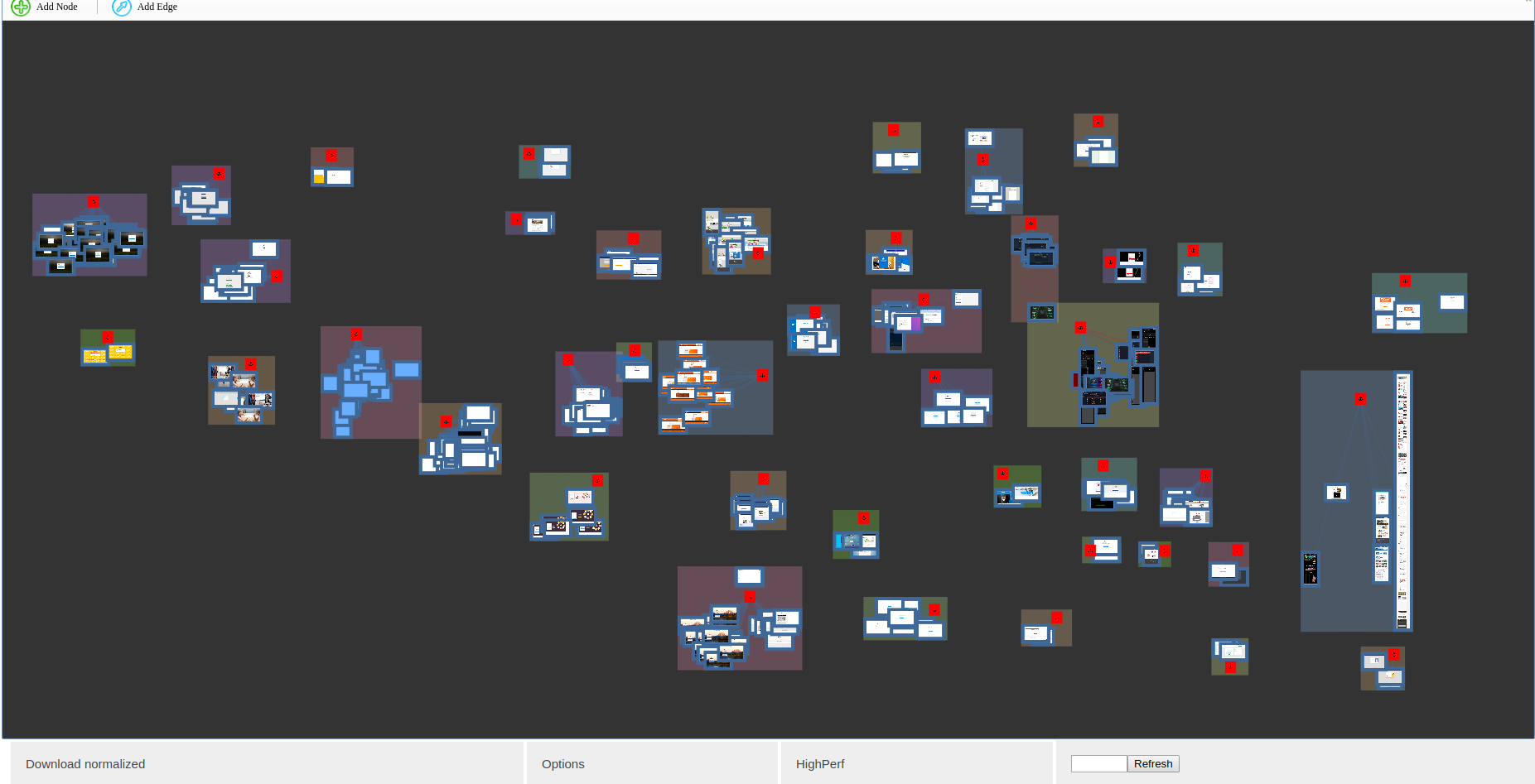}
    \caption{VISJS Classificator labelling interface after classification. All pictures are clustered in their own cluster. (Physics disabled)}
    \label{fig:afterlabeling}
   \end{subfigure}
  
    \begin{subfigure}[b]{\textwidth}
    \includegraphics[width=\linewidth]{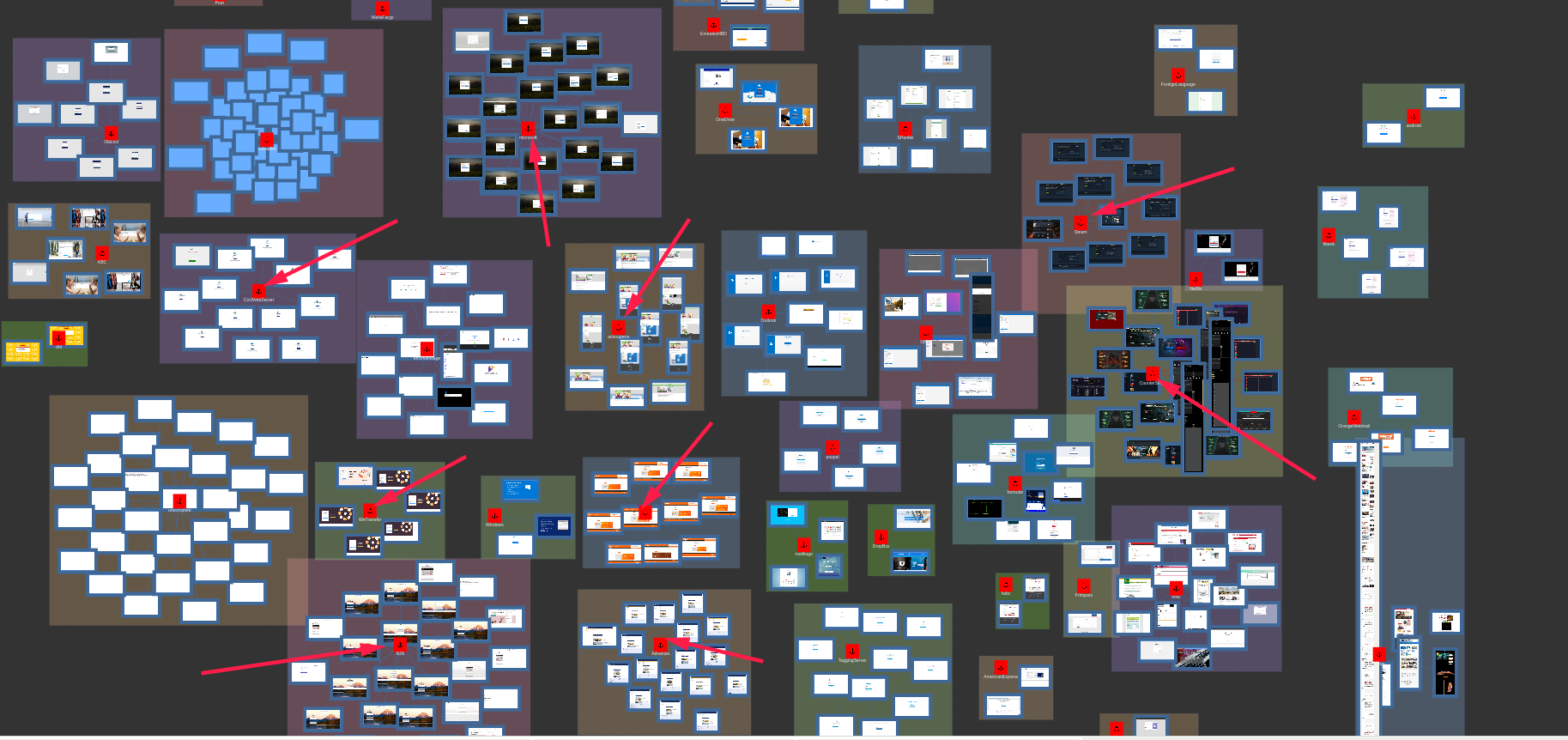}
    \caption{Few examples of clusters found in the first ten minutes of manual clustering, that were kept until the end. (Physics enabled)}
    \label{fig:withphysics}
  \end{subfigure}
  
  \caption{User interfaces after labeling}
  \label{fig:user-interfaces}
\end{figure}

\clearpage
\subsubsection{Feedback}

We asked few questions to the participants with a form to complete after the experiments. Answers were anonymous and collected few hours after their participation to experiments. Results are presented in Figure \ref{fig:feedback}, page \pageref{fig:feedback}.\\ 
Overall, there is a gap between those who liked VisJS and those who didn't. Standard labeling tools are still above our graph-based tool in terms of easiness of use, speed and entertainment.\\
Future improvements may change these conclusions.

\begin{table}
\centering
\begin{tabular}{llcl}
\textbf{Question} & \textbf{Subquestion} & \textbf{Mean score (5 = positive)} &   \\
How satisfied were you of this experiment ?			&& 4&   \\
How did you evaluated your speed on DataTurks ?  		&& 3&   \\
How did you evaluated your speed on VisJS ?  			&& 3 (high volatily)&   \\
What would you say about DataTurks ?					&& / &   \\
-                 & It's easy to add labels ?        & 2&   \\
-                 & It's easy to use (general) ?     & 4.5&   \\
-                 & It's fast ?                    	&  4&   \\
-                 & It allows a consistent labeling ? &  4&   \\
-                 & It allows to have a good quality labeling ?& 3 (high volatily)&   \\
-                 & That was a positive experience ? & 4.5&   \\
-                 & That was fun ?                   & 4.5&   \\
What would you say about VisJS ?						&&  / &   \\
-                 & It's easy to add labels ?        & 3 (high volatily)&   \\
-                 & It's easy to use (general) ?     & 3 (high volatily)&   \\
-                 & It's fast ?                    	& 2.5&   \\
-                 & It allows to have a good quality labeling ?& 4.5&   \\
-                 & That was a positive experience ? & 4 (high volatily)&   \\
-                 & That was fun ?                   & 3.5&   \\
Which tool did you prefer ? &  &       50\%/50\%              &   \\
\end{tabular}
  \caption{Participants feedback about the tools and he experiment.}
  \label{fig:feedback}
\end{table}

\section{Conclusion}

The main contribution of this paper is the introduction of a new tool to manually classify pictures with prior unknown labels, nor number of labels.
The tool is provided as an Open-Source software, available at \href{https://github.com/Vincent-CIRCL/visjs_classificator}{github.com/Vincent-CIRCL/visjs\_classificator} \cite{vincent-circlClassificatorPicturesMatching2019a}

\bibliographystyle{paper-ressources/IEEEtran}
\bibliography{./carl-hauser.bib}

\begin{thebibliography}{1}
\providecommand{\url}[1]{#1}
\csname url@samestyle\endcsname
\providecommand{\newblock}{\relax}
\providecommand{\bibinfo}[2]{#2}
\providecommand{\BIBentrySTDinterwordspacing}{\spaceskip=0pt\relax}
\providecommand{\BIBentryALTinterwordstretchfactor}{4}
\providecommand{\BIBentryALTinterwordspacing}{\spaceskip=\fontdimen2\font plus
\BIBentryALTinterwordstretchfactor\fontdimen3\font minus
  \fontdimen4\font\relax}
\providecommand{\BIBforeignlanguage}[2]{{%
\expandafter\ifx\csname l@#1\endcsname\relax
\typeout{** WARNING: IEEEtran.bst: No hyphenation pattern has been}%
\typeout{** loaded for the language `#1'. Using the pattern for}%
\typeout{** the default language instead.}%
\else
\language=\csname l@#1\endcsname
\fi
#2}}
\providecommand{\BIBdecl}{\relax}
\BIBdecl

\bibitem{mokaddemAILDesignImplementation2018}
S.~Mokaddem, G.~Wagener, and A.~Dulaunoy, ``{{AIL}} - {{The}} design and
  implementation of an {{Analysis Information Leak}} framework,'' in \emph{2018
  {{IEEE International Conference}} on {{Big Data}} ({{Big Data}})}, pp.
  5049--5057.

\bibitem{vincent-circlClassificatorPicturesMatching2019a}
\BIBentryALTinterwordspacing
Vincent-CIRCL, ``Classificator for pictures matching and clustering. {{Fast}}
  and visual.: {{Vincent}}-{{CIRCL}}/visjs\_classificator.'' [Online].
  Available: \url{https://github.com/Vincent-CIRCL/visjs_classificator}
\BIBentrySTDinterwordspacing

\end{thebibliography}

\end{document}